\documentstyle[aps,prb]{revtex}
\begin{document}
\pagestyle{plain}
\newcommand{\D}{\displaystyle}
\title{\bf Combined potential and spin impurity scattering in cuprates} 
\author{G. Hara\'n$^{\star}$ and A.D.S. Nagi$^{\dag}$}
\address{$^{\star}$Institute of Physics, Politechnika
Wroc{\l}awska, Wybrze\.ze Wyspia\'nskiego 27, 50-370 Wroc{\l}aw,
Poland}
\address{$^{\dag}$Department of Physics, University of Waterloo,
Waterloo, Ontario, Canada, N2L 3G1}
\date{11 September 2000}
\maketitle

\begin{abstract}
We present a theory of combined nonmagnetic and magnetic impurity scattering 
in anisotropic superconductors accounting for the momentum-dependent 
impurity potential. Applying the model to the d-wave superconducting state,  
we obtain a quantitative agreement with the initial suppression of the critical 
temperature due to Zn and Ni substitutions as well as electron irradiation defects 
in the cuprates. We suggest, that the unequal pair-breaking  
effect of Zn and Ni may be related to a different nature of the magnetic 
moments induced by these impurities.  
\end{abstract}
\vspace*{0.5cm}

Impurities offer a useful experimental probe of the fundamental properties of 
high temperature superconductors. Controlled substitutions of 3d transition metals 
(Zn, Ni) provide indirect information on the nature of the pairing mechanism since  
they affect the physical properties of both the superconducting and normal state. 
The nominally nonmagnetic Zn ($3d^{10}$, $S=0$) as well as magnetic Ni ($3d^8$, $S=1$) 
atoms reside in the magnetically active in-plane Cu ($3d^9$, $S=1/2$) sites. 
Macroscopic susceptibility and NMR measurements \cite{S3,S3a,S1,S2,S2a} 
of $YBa_2Cu_3O_{7-\delta}$ (Y-123) superconductor indicate the impurity-induced 
magnetic moments of $0.86\mu_B$ for Zn, and $1.9\mu_B$ for Ni in the underdoped 
compound, which decrease 
with hole doping to $0.36\mu_B/Zn$ and $1.6\mu_B/Ni$ in the optimally doped 
system. In $La_{2-x}Sr_xCuO_4$ (La-214) the same impurity substitutions lead to 
magnetic properties corresponding to local magnetic moments of $1.0\mu_B/Zn$ and 
$0.6\mu_B/Ni$. \cite{S2b} Magnetism generated by Zn and Ni 
atoms is, however, different in nature. Whereas, the Zn-induced moments reside 
in the vicinity of the nonmagnetic impurity on the neighboring Cu sites, 
\cite{S1,S2,S2a,S4} the Ni substitution yields the magnetic moments which partially 
screen $S=1$ impurity spin and result in a more localized $S=1/2$ moment. 
\cite{S2a,S4} 
In the cuprates, where the strong Coulomb interactions are present, 
a creation of a magnetic moment around any lattice defect cannot be excluded.  
Particularly, a removal of the oxygen atom from the $CuO_2$ plane, \cite{O1,O2,O3,O4} 
where the Cu ($3d$) spins and O ($2p$) holes are strongly coupled, should  
give rise to a local uncompensated magnetic moment, that is it should develop a 
magnetically active oxygen vacancy defect.   

A theoretical analysis of the impurity influence on the d-wave superconducting 
state based on a standard isotropic (s-wave) scattering approach indicates a 
significantly stronger suppression of superconductivity than observed in the 
cuprates. \cite{5} This discrepancy can be reconciled within a simple model 
of anisotropic impurity scattering \cite{1,2} which takes also the non s-wave 
scattering channels into account. Evaluation of the critical temperature \cite{1}  
and the upper critical field \cite{M1,M2} dependences on the scattering 
rate showed their robustness to impurity scattering in the d-wave channel and 
led to a quantitative agreement with the experimental data. A strong support to  
these early theoretical predictions is provided by a recent experiment \cite{Lin} 
with the electron irradiation created oxygen defects in Y-123 which confirms 
the extended nature of the scattering centers and the ability of the model to 
account for both the $T_c$ and $H_{c_{2}}$ suppression by the disorder. 
Despite its preliminary success this approach neglects the magnetic scattering 
processes which however should be present at least for Zn and Ni impurity  
substitutions. Obviously, the spin-flip scattering leads to an additional 
destruction of a singlet d-wave superconducting state and, if included in 
our model, could possibly shift its predictions beyond the experimental range.  
In this paper, we discuss this issue in more detail by generalizing the effective 
anisotropic impurity scattering \cite{1,2} to a combined potential and magnetic 
scattering. We analyze the initial $T_c$ suppression in $YBa_2Cu_3O_{7-\delta}$, 
$La_{2-x}Sr_xCuO_4$, $Bi_2Sr_2CaCu_2O_{8+\delta}$ (Bi-2212) 
compounds induced by Zn \cite{Zn1,Zn2,Zn3,Zn4,Zn5,Zn7,Zn8} and Ni \cite{Zn5,Ni1} 
substitution as well as by the in-plane oxygen vacancy defect created by  
the electron irradiation. \cite{O1,O2,O3,O4}  
Discussing the $T_c$ alone cannot give a final test of the impurity scattering 
model. Its applicability can be ultimately verified by a thorough analysis  
of the thermodynamic and transport properties as has been described for isotropic 
scattering in Refs. 26-28. Such a comprehensive approach is beyond the scope of 
this brief report in which we focus on a possible role of the extended magnetic 
scattering in the suppression of the critical temperature. An elaborated 
study of the presented model will be given in a separate paper. 

We consider randomly distributed impurities at low concentration 
interacting with conduction electrons through a potential   
$u({\bf k},{\bf k}')=v({\bf k},{\bf k}')+J({\bf k},{\bf k}'){\bf\bar{S}\bar{\sigma}}$, 
where ${\bf\bar{S}}$ is a classical spin representing the impurity and 
${\bf\bar{\sigma}}$ 
is the electron spin density. The anisotropic superconducting state is   
determined by the order parameter $\Delta({\bf k})=\Delta e({\bf k})$, 
where $e\left({\bf k}\right)$ is a momentum-dependent function.  
We normalize $e({\bf k})$ in this way that $\left<e^{2}\right>=1$, 
where $<...>=\int_{FS}dS_k n({\bf k})(...)$
denotes the average value over the Fermi surface (FS) momenta,
and $n({\bf k})$ is the angle resolved FS density
of states normalized to unity, i.e. $\int_{FS}dS_k n({\bf k})=1$.
For the $(d_{x^2-y^2}+s)$-wave state \mbox{$e({\bf k})
=\left(cos2\phi +s\right)/\left<\left(cos2\phi + s\right)^2
\right>^{1/2}$} in a polar angle notation where the $d_{x^2-y^2}$-wave 
state corresponds to $s=0$. Taking the electron-impurity scattering within 
the Born approximation and neglecting the impurity-impurity
interaction, \cite{3} the diagonal and off-diagonal corrections to the Green's   
function averaged over the impurity positions and spin directions read \cite{3,4} 

\begin{equation}
\label{e1}
\tilde{\omega}\left({\bf k}\right)=\omega+\pi n_iN_0
\int_{FS}dS_{\bf k'}n\left({\bf k}'\right)\tilde{\omega}\left({\bf k}'\right) 
\frac{\D\left[v^2\left({\bf k},{\bf k}'\right)+
J^2\left({\bf k},{\bf k}'\right)S\left(S+1\right)\right]}
{\left[\D\tilde{\omega}^2\left({\bf k}'\right)+
\tilde{\Delta}^2\left({\bf k}'\right)\right]^{1/2}}
\end{equation}

\begin{equation}
\label{e2}
\tilde{\Delta}\left({\bf k}\right)=\Delta\left({\bf k}\right)+ 
\pi n_iN_0\int_{FS}dS_{\bf k'}n\left({\bf k}'\right)\tilde{\Delta}\left({\bf k}'\right)
\frac{\D\left[v^2\left({\bf k},{\bf k}'\right)-
J^2\left({\bf k},{\bf k}'\right)S\left(S+1\right)\right]}
{\left[\D\tilde{\omega}^2\left({\bf k}'\right)+
\tilde{\Delta}^2\left({\bf k}'\right)\right]^{1/2}}
\end{equation}

\noindent
where $\omega=\pi T(2n+1)$ is the Matsubara frequency 
(T is temperature and n is an integer number),  
$N_0$ is the single-spin density of states at the 
Fermi level, $n_i$ is the impurity (defect) concentration, and all the wave vectors 
are restricted to the Fermi surface. We generalize our  
earlier model \cite{1,2} to include magnetic scattering and assume that the 
momentum-dependent potential terms in Eqs. (\ref{e1}), (\ref{e2}) are separable 
and given by 

\begin{eqnarray}
\label{e3}
v^2\left({\bf k},{\bf k}'\right)&=&v_0^2+v_1^2
f\left({\bf k}\right)f\left({\bf k}'\right)\\
J^2\left({\bf k},{\bf k}'\right)&=&J_0^2+J_1^2
g\left({\bf k}\right)g\left({\bf k}'\right)
\end{eqnarray}

\noindent
where $v_0$ ($v_1$), $J_0$ ($J_1$) are isotropic (anisotropic) scattering amplitudes 
and $f\left({\bf k}\right)$, $g\left({\bf k}\right)$ are the momentum-dependent 
anisotropy functions in the nonmagnetic and magnetic scattering channel, respectively. 

We assume that the overall scattering rate is determined by the isotropic
components and to have it well defined impose the constraints

\begin{eqnarray}
\label{e4}
v_1^2\leq v_0^2,\;\;\;\;&\;\;\;\;\left<f\right>=0,\;\;\;\;&\;\;\;\;\left<f^2\right>=1\\
J_1^2\leq J_0^2,\;\;\;\;&\;\;\;\;\left<g\right>=0,\;\;\;\;&\;\;\;\;\left<g^2\right>=1
\end{eqnarray}

\noindent
Note, that such a choice of the anisotropy does not change the normal 
state properties as the diagonal part of the self-energy in the limit of 
$\Delta\rightarrow 0$ is given by 
$\tilde{\omega}\left({\bf k}\right)_{\Delta=0}=\omega+\pi n_iN_0
\left(v_0^2+J_0^2S\left(S+1\right)\right)sgn\left(\omega\right)$ and depends only 
on the isotropic scattering rates.   

Inclusion of a momentum-dependent model impurity potential in the magnetic 
channel is motivated by the existence of extended magnetic moments associated with  
the impurity, \cite{S3,S3a,S1,S2,S2a,S2b,S4} and the anisotropy of the nonmagnetic  
scattering channel simulates the anisotropy of the crystal lattice.      
For Ni impurity which leads to a screened localized magnetic  
moment \cite{S2a,S4} we expect $J_1/J_0\ll 1$, while for broadly distributed over Cu 
sites Zn-induced magnetic moment \cite{S1,S2,S2a,S4} both isotropic and anisotropic 
scattering channels should be comparable in magnitude, i.e. $J_1/J_0\sim 1$.  
Contrary to the NMR measurements \cite{S3a,S1,S2,S2a} muon spin rotation ($\mu SR$) 
experiments \cite{S5} report no evidence for additional Zn-induced moments 
in underdoped Y-123. Yet, one cannot exclude the possibility that the 
impurity-generated magnetic moments exist but are weakly coupled to the 
spin system of the $CuO_2$ planes. Therefore, the following assumption about 
the relative amount of magnetic scattering seems reasonable:  
$J_1/v_1< 1$, $J_0/v_0< 1$. 

Taking a separable pair potential 
$V\left({\bf k},{\bf k'}\right)=-V_0e\left({\bf k}\right)
e\left({\bf k'}\right)$, $V_0>0$, and following a standard procedure \cite{1,4} 
we obtain the critical temperature, $T_c$, equation    

\begin{equation}
\label{e5}
\begin{array}{l}
\D\ln\frac{\D T_c}{\D T_{c_{0}}}=\left(1-\left<e\right>^2\right)
\left[\psi\left(\frac{\D 1}{\D 2}\right)-
\psi\left(\frac{\D 1}{\D 2}+\frac{\D\Gamma_0+G_0}{\D 2\pi T_c}\right)\right]\\ 
\\
+\left<e\right>^2\left[\psi\left(\frac{\D 1}{\D 2}\right)
-\psi\left(\frac{\D 1}{\D 2}+\frac{\D 2G_0}{\D 2\pi T_c}\right)\right] 
+S_1+S_2
\end{array}
\end{equation}

\noindent
where $T_{c_{0}}$ is the critical temperature in the absence of impurities, 
$\psi\left(z\right)$ is the digamma function,   

\begin{displaymath}
\begin{array}{l}
\D S_1=2\pi T_c\sum_{\omega>0}\frac{\D\left<ef\right>}
{\D\left(\omega+\Gamma_0+G_0\right)}\frac{\D\Gamma_1\left<ef\right>
\left(\omega+\Gamma_0+G_0+G_1\right)-\Gamma_1G_1\left<eg\right>\left<fg\right>} 
{\D\left(\omega+\Gamma_0+G_0+G_1\right)\left(\omega+\Gamma_0+G_0-\Gamma_1\right)
+G_1\Gamma_1\left<fg\right>^2}\\
\\
\D S_2=-2\pi T_c\sum_{\omega>0}\frac{\D\left<eg\right>}
{\D\left(\omega+\Gamma_0+G_0\right)}\frac{\D G_1\left<eg\right>
\left(\omega+\Gamma_0+G_0-\Gamma_1\right)+\Gamma_1G_1\left<ef\right>\left<fg\right>}
{\D\left(\omega+\Gamma_0+G_0+G_1\right)\left(\omega+\Gamma_0+G_0-\Gamma_1\right)
+G_1\Gamma_1\left<fg\right>^2}
\end{array}
\end{displaymath}

\noindent
and $\Gamma_0=\pi n_iN_0v_0^2$, $\Gamma_1=\pi n_iN_0v_1^2$,    
$G_0=\pi n_iN_0J_0^2S\left(S+1\right)$, and $G_1=\pi n_iN_0J_1^2S\left(S+1\right)$ 
are the intrinsic scattering rates. Given strongly coupled spin and charge  
dynamics in the copper-oxygen planes the impurity potential encountered 
by the electron in the spin scattering channel should follow the one in the 
nonmagnetic channel, that is $g\left({\bf k}\right)=\pm f\left({\bf k}\right)$,   
which simplifies Eq. (\ref{e5}) to 

\begin{equation}
\label{e7}
\begin{array}{l}
\D\ln\frac{T_c}{T_{c_{0}}}=\left(1-\left<e\right>^2-\left<ef\right>^2\right) 
\left[\psi\left(\frac{\D 1}{\D 2}\right)-
\psi\left(\frac{\D 1}{\D 2}+\frac{\D\Gamma_0+G_0}{2\pi T_c}\right)\right]\\
\\
+\D\left<ef\right>^2\left[\psi\left(\frac{\D 1}{\D 2}\right)-\psi\left(
\frac{\D 1}{\D 2}+\frac{\Gamma_0+G_0+G_1-\Gamma_1}{2\pi T_c}\right)\right] 
+\left<e\right>^2\left[\psi\left(\frac{\D 1}{\D 2}\right)-
\psi\left(\frac{\D 1}{\D 2}+\frac{\D 2G_0}{2\pi T_c}\right)\right]
\end{array}
\end{equation}

\noindent
Comparing the above equation with the one for nonmagnetic scattering 
(Eq. (24) of Ref. 13) we notice that there is only one additional term in  
Eq. (\ref{e7}) reflecting an extra $T_c$ suppression due to the spin-flip 
scattering determined by the exchange rate $2G_0$. The remaining terms of  
Eq. (\ref{e7}) correspond to the ones obtained for potential scattering \cite{1} 
with some new effective scattering rates being combinations of the scattering 
rates in the magnetic and nonmagnetic channels: $\bar{\Gamma}_0=\Gamma_0+G_0$,  
$\bar{\Gamma}_1=\Gamma_1-G_1$. We note, that the anisotropy of the impurity 
potential has no effect on the $T_c$ when $\bar{\Gamma}_1=0$, i.e. when 
$\Gamma_1=G_1$, which should not be the case of the cuprates, where most 
probably the exchange interaction between the impurity spin and the 
Cu-O spin system is small, $G_1/\Gamma_1< 1$. The role of anisotropy in 
the impurity potential is concurrently expressed by the dimensionless 
parameter $\left<ef\right>^2$ ($0\leq\left<ef\right>^2\leq 1$) representing  
the interplay between the pair potential and the anisotropic part of the 
impurity potential. Again, for $\left<ef\right>^2=0$ we obtain a regular $T_c$ 
suppression due to the isotropic combined nonmagnetic and magnetic impurity 
scattering \cite{7} which in the case of a conventional s-wave superconductor 
reduces to the well known relationship. \cite{3,4} 

In the quantitative analysis of the experimental data we consider  
the $T_c$ reduction in the limit of low impurity concentration
$n_i\rightarrow 0$ which is given by the initial slope

\begin{equation}
\label{e8}
\D\frac{\D dT_c}{\D d\bar{\Gamma}_0}=-\frac{\D\pi}{\D 4} 
\left[\chi+\left(1-\chi\right)\phi_M-\phi_A\right]
\end{equation}

\noindent
where for the sake of brevity we have introduced $\chi=1-\left<e\right>^2$, 
$\phi_A=\left<ef\right>^2\bar{\Gamma}_1/\bar{\Gamma}_0$ (anisotropy factor), 
and $\phi_M=2G_0/\bar{\Gamma}_0$ (magnetic factor). The isotropic part of 
the impurity potential $\bar{\Gamma}_0$ determines the residual resistivity    
at the zero frequency \cite{1}   
$\rho_0\simeq 10.18\times 10^{-2}\;\frac{\D 8\pi^2}{\D\omega^2_{pl}}
T_{c_0}\left(\frac{\D\bar{\Gamma}_0}{\D 2\pi T_{c_0}}\right)\mu\Omega cm$,    
where $\omega_{pl}$ is the in-plane plasma frequency in eV.  
We fix $\omega_{pl}$ within the range \cite{5} 
$1.1 eV\leq\omega_{pl}\leq 1.4eV$ for Y-123, and estimate it from the 
zero temperature penetration depth measurements \cite{14,19,20} as 
$0.84eV$ and $0.9eV$ for La-214 and Bi-2212, respectively. There is still no 
complete experimental agreement on the impurity-induced resistivity in the cuprates. 
\cite{Zn1,Zn2,Zn5,Zn8,Ni1} However, some results \cite{Zn1,Zn5} suggest that 
Zn and Ni lead to a comparable effect in the normal state, which would mean 
that $\bar{\Gamma}_0\left(Zn\right)\sim \bar{\Gamma}_0\left(Ni\right)$. 
The residual resistivity $\rho_0$ is used to express the initial $T_c$ suppression in 
the following form convenient for the discussion of the experimental data

\begin{equation}
\label{e9}
\D\frac{dT_c}{d\rho_0}\simeq -0.614\:\omega^2_{pl}
\left[\chi+\left(1-\chi\right)\phi_M-\phi_A\right]K/\mu\Omega cm
\end{equation}

\noindent 
We find that in the d-wave state ($\chi=1$) the initial $T_c$ suppression  
(Eq. (\ref{e9})) is determined only by the dimensionless anisotropy factor $\phi_A$.  
The influence of the spin-flip scattering is present through 
the renormalized scattering rates in the isotropic and anisotropic 
scattering channels: $\bar{\Gamma}_0$ and $\bar{\Gamma}_1$.  
However, the magnetic factor $\phi_M$ has no effect on the critical 
temperature and we may say that the d-wave system becomes insensitive 
to magnetic scattering as the solutions of Eq. (\ref{e7}) correspond to 
those for nonmagnetic scattering \cite{1} with renormalized parameters.  
In Tabs. I-III we give the anisotropy factors $\phi_A$ needed 
to account for the experimental data of Zn, Ni and oxygen vacancy-induced 
$T_c$ suppression in Y-123, La-214 and Bi-2212 compounds. \cite{proof} 
Worth noting are generally significantly lower values of $\phi_A$ for Zn substitution 
than for Ni. This feature can be attributed to a different nature of the 
magnetic moments associated with these impurities. While the Zn-induced moment 
is broadly distributed over neighboring Cu sites ($G_1/G_0\sim 1$), 
the one created by Ni is screened on a larger distance and is more localized 
($G_1/G_0\ll 1$). Therefore, for comparable amounts of isotropic scattering  
by these impurities, $\bar{\Gamma}_0=\Gamma_0+G_0$, the anisotropy factor $\phi_A\sim 
\left(\Gamma_1-G_1\right)/\left(\Gamma_0+G_0\right)$ of Ni scattering potential,  
$\sim\Gamma_1/\left(\Gamma_0+G_0\right)$, should exceed the one of Zn impurity 
$\sim\left(\Gamma_1-G_0\right)/\left(\Gamma_0+G_0\right)$.  
Given large $\phi_A$ values for the oxygen vacancy we may also infer that the 
magnetic moments formed around this defect are rather localized or the anisotropic 
scattering takes place in the d-wave channel, i.e. $\left<ef\right>^2\sim 1$. 
Still, to decide definitely about the role of spin scattering in high $T_c$ 
materials with simple defects a quantitative estimate of the ratios of the 
magnetic to nonmagnetic scattering rates $G_0/\Gamma_0$ and $G_1/\Gamma_1$ is 
required. Of some help here may be a generalization of the $H_{c_{2}}$ critical 
field analysis to magnetic impurity scattering. So far, an application of the  
nonmagnetic scattering expression for the reduced $H_{c_{2}}$ slope \cite{M1,M2} 
in the interpretation of the critical field of the electron irradiated Y-123 has 
confirmed the anisotropy parameters of the model determined from $T_c$ suppression 
in this compound. \cite{Lin} 

The orthorhombic distortion of the crystal lattice in the Y-123 superconductor leads 
to the (d+s)-wave superconductivity in this compound. \cite{21} The photoemission,  
tunneling and thermodynamic measurements put an upper bound of $\sim 10\%$ on the 
relative size of the s-wave component in the (d+s)-wave superconducting state. 
\cite{21} If we define a per cent fraction $\Delta_{s\%}$ of the $s$-wave component 
as a ratio of a minimal gap to its maximum value, then the condition $\Delta_{s\%}=
10\%$ is equivalent to $\chi=0.976$ ($\left<e\right>^2=0.0241$). It is a matter of 
a straightforward calculation to show, that $\phi_A$ for a given superconducting 
state is related to the one of the d-wave state $\phi_A^0$ through 
$\phi_A=\phi_A^0-\left(1-\chi\right)\left(1-\phi_M\right)$. In what follows, for 
$\chi=0.976$ we get merely a slight change of the anisotropy factor 
$\phi_A=\phi_A^0-0.024\left(1-\phi_M\right)$ which for different amounts of the 
magnetic scattering rate reads: $\phi_A=\phi_A^0-0.024$ for $\phi_M=0$ 
($G_0=0$); $\phi_A=\phi_A^0-0.008$ for $\phi_M=0.666$ ($G_0/\Gamma_0=0.5$); 
$\phi_A=\phi_A^0$ for $\phi_M=1$ ($G_0/\Gamma_0=1$). Our approach reduces  
to the isotropic impurity scattering approximation \cite{7} in the limit of $\phi_A=0$.  
Therefore, given the nonzero $\phi_A$ values in Tabs. I-III and taking the above  
corrections for the (d+s)-wave state into account, we note that the experimental 
data cannot be explained by applying the isotropic impurity scattering approach 
to the (d+s)-wave superconducting state with an appropriate (for the cuprates)  
amount of the s-wave component.  
  
Concluding, we find that even in the presence of the  
spin-flip processes the scenario of anisotropic impurity scattering accounts 
quantitatively for the experimentally observed $T_c$ suppression due to Zn, 
Ni and oxygen vacancy scattering. Moreover, a different partition of the magnetic 
scattering rates into isotropic and anisotropic scattering channels can 
be used in the interpretation of the weak Ni-induced and relatively strong 
Zn-induced pair-breaking. Our result also shows that the isotropic impurity  
scattering approximation does not explain quantitatively the initial 
suppression of the critical temperature in the cuprates. 
Finally, we note that a significant sample-dependence of the discussed 
data with the discrepancy even as high as $50\%$ calls for further more  
controlled measurements of the impurity-induced $T_c$ suppression in 
the high temperature superconductors.\\ 

The authors would like to thank Professor Peter Fulde for helpful comments 
and hospitality at Max-Planck-Institut f\"ur Physik komplexer Systeme in 
Dresden where the manuscript was prepared.

\newpage
\begin{table}
\caption{{\bf Y-123}: $\phi_A$ anisotropy factor of the impurity potential reproducing  
the $T_c$ suppression in the Y-123 compound within the d-wave superconductivity 
scenario, i.e. $\chi=1$. ($1.1eV\leq\omega_{pl}\leq 1.4eV$;   
$ov.$=overdoped, $op.$=optimally doped, $un.$=underdoped)}  
\vspace*{0.2cm}
\begin{tabular}{||c|c|c|c||}
$defect$&$sample$&$\left(\frac{\D dT_c}{\D d\rho_0}\right)_{exp}\;\left[K/\mu\Omega cm\right]$
&$anisotropy\;factor:\;\phi_A$\\
&&&\\ \hline
&$(ov.)\;single\;crystal^{18} $&$-0.674$&$ 0.093\leq\phi_A\leq 0.440$\\
&&&\\
$Zn$&$(ov.)\;single\;crystal^{19,22}$&$-0.57$&$0.233\leq\phi_A\leq 0.526$\\
$impurity$&&&\\
&$(op.)\;thin\;film^{23}$&$-0.241$&$0.676\leq\phi_A\leq 0.780$\\
&&&\\
&$(op.)\;thin\;film^{24}$&$-0.520$&$0.300\leq\phi_A\leq 0.568$\\ \hline
&&&\\
$Ni$&$(ov.)\;ceramic\;sample^{22}$&$-0.333$&$0.552\leq\phi_A\leq 0.723$\\
$impurity$&&&\\
&$(ov.)\;film^{25}$&$from\;(-0.063)\;to\;(-0.044)$&$0.915\leq\phi_A\leq 0.963$\\ \hline
&&&\\
$oxygen$&$(ov.)\;single\;crystal^{9}$&$-0.30\pm 0.04$&$0.542\leq\phi_A\leq 0.784$\\
$vacancy$&&&\\
&$(ov.)\;film^{8}$&$-0.187$&$0.748\leq\phi_A\leq 0.845$
\end{tabular}
\end{table}

\begin{table}
\caption{{\bf La-214}: $\phi_A$ anisotropy factor of the impurity potential 
reproducing  
the $T_c$ suppression in the La-214 compound within the d-wave superconductivity 
scenario, i.e. $\chi=1$. ($\omega_{pl}=0.84eV$; notation as in Tab. I.)}  
\vspace*{0.2cm}
\begin{tabular}{||c|c|c|c||}
$defect$&$sample$&$\left(\frac{\D dT_c}{\D d\rho_0}\right)_{exp}\;
\left[K/\mu\Omega cm\right]$
&$anisotropy\;factor:\;\phi_A$\\
&&&\\ \hline
$Zn$&$(ov.)\;single\;crystal^{19}$&$-0.37$&$0.146$\\
$impurity$&&&\\
&$(un.)\;film^{20,21}$&$-0.233$&$0.462$\\ \hline
&&&\\
$oxygen$&$(un.)\;film^{10}$&$-0.127$&$0.707$\\
$vacancy$&&&
\end{tabular}
\end{table}

\begin{table}
\caption{{\bf Bi-2212}: $\phi_A$ anisotropy factor of the impurity potential 
reproducing  
the $T_c$ suppression in the Bi-2212 compound within the d-wave superconductivity
scenario, i.e. $\chi=1$. ($\omega_{pl}=0.9eV$; notation as in Tab. I.)}
\vspace*{0.2cm}
\begin{tabular}{||c|c|c|c||}
$defect$&$sample$&$\left(\frac{\D dT_c}{\D d\rho_0}\right)_{exp}\;
\left[K/\mu\Omega cm\right]$
&$anisotropy\;factor:\;\phi_A$\\
&&&\\ \hline
$oxygen$&$(ov.)\;single\;crystal^{11}$&$-0.28$&$0.437$\\
$vacancy$&&&
\end{tabular}
\end{table}


\begin{references}
\bibitem{proof} Except for Ni effect in overdoped Y-123 films (Ref. 25) the 
anisotropy factor of the d-wave state (Tabs. I-III) can be chosen lower than 
a cylindrical Fermi surface average of the absolute value of the order parameter 
$\left<\left|e\right|\right>\simeq 0.81$.
\end{references}
\end{document}